# A CYBER-PHYSICAL DIGITAL TWIN APPROACH TO REPLICATING REALISTIC MULTI-STAGE CYBERATTACKS ON SMART GRIDS

Ömer Sen
Fraunhofer FIT – Germany

oemer.sen@fit.fraunhofer.de

Nathalie Bleser
IAEW at RWTH Aachen - Germany

nathalie.bleser@rwth-aachen.de

Martin Henze
RWTH Aachen University and Fraunhofer FKIE - Germany
henze@cs.rwth-aachen.de

Andreas Ulbig
IAEW at RWTH Aachen - Germany

a.ulbig@iaew.rwth-aachen.de

## ABSTRACT

*The integration of information and communication technology in distribution grids presents opportunities for active grid operation management, but also increases the need for security against power outages and cyberattacks. This paper examines the impact of cyberattacks on smart grids by replicating the power grid in a secure laboratory environment as a cyber-physical digital twin. A simulation is used to study communication infrastructures for secure operation of smart grids. The cyber-physical digital twin approach combines communication network emulation and power grid simulation in a common modular environment, and is demonstrated through laboratory tests and attack replications.*

## INTRODUCTION

The integration of Information and Communications Technology (ICT) into distribution grids is resulting in an increasing digitalization of the grid. This trend offers enhanced transparency through real-time monitoring and active control of individual grid assets, and presents new possibilities for active operational management at the distribution grid level, particularly in terms of optimizing the use of flexibility resources provided by distributed generation and consumers in a market- and grid-related manner.

The incorporation of ICT in grid operations also requires the fulfilment of fundamental requirements for the cyber-resilience and cybersecurity of the overall system [1]. Failures in the ICT domain can have an impact on the security of supply, so measures are necessary to prevent cyberattacks and address the increased attack surface. To develop, validate, and test such measures, access to communications data from attack patterns is necessary. However, such data is not publicly available [2], mainly due to security and confidentiality concerns.

The development of new operational management concepts and the investigation of domain-specific cybersecurity measures both require tailored development environments that are scalable, realistic, and flexible. To study the cybersecurity of the power grid and generate attack data without compromising the grid itself, a feasible approach is necessary.

In this paper, we propose a Cyber-Physical Digital Twin (CPDT) approach that enables the investigation of power grid cybersecurity without compromising grid security by generating attack data through cyberattack replications in Smart Grid (SG). Our approach is built on a co-simulation framework integrated into a laboratory environment with the ability to replicate SG in its process and ICT layers. Our specific contributions in this paper are as follows:

- We present the current state of the art in digital twin-based test and investigation environments and highlight the issue of attack data generation.
- We describe the overall proposed approach, including the co-simulation framework, the laboratory environment, and the reference grid.
- We demonstrate and discuss the cybersecurity investigation capabilities of our proposed approach through a case study of cyberattacks.

This paper presents our ongoing work on a framework for co-simulating cyber-physical energy systems. In the co-simulation section, we present the goals and added values of the environment. The section regarding CPDT gives an overview of the environment and its main components, including the integration of operational concepts for flexibility management and cyberattack replication. The main focus of the paper is on the presentation of our proposed CPDT approach using a co-simulation environment, which we demonstrate through a case study investigation in the result section.

## CO-SIMULATION FOR SMART GRIDS

This section covers (i) the structure and considerations for co-simulation approaches in SGs, (ii) research on replicating SGs and cybersecurity using co-simulation, and (iii) concludes with a problem statement outlining the research scope.

### Smart Grid Simulation

The overall communication infrastructure of a Distribution System Operator (DSO) and the components to be represented within the simulation environment is based on the Purdue Model [3]. The SG architecture based on the Purdue Model is described by dividing the components into the domains of Energy Technology (ET), Operational Technology (OT), and Information Technology (IT). Within this work, we define the requirements for our co-simulation environment such that it must map the mutual dependencies and interactions between the domains of ET, OT, and IT as a cyber-physical system. The co-simulation must consider the scheduling of discrete-time stepped simulation of power grids and emulation of communication networks using a discrete time stepped scheduling method with a small step size. It must also map the behavior and reaction of the grid operator based on changes in the grid condition. The design of the co-simulation is largely influenced by the mapping of the individual components of the IT and OT domains and their realistic communication behavior. The goal is to simulate cyberattacks by systematically mapping such events for individual components or groups of components. In terms of threats to reliable power supply, both direct attacks on OT systems and multi-stage cyberattacks that propagate laterally from vulnerabilities accessible through public networks must be considered. To realistically model attacks and the exploitation of vulnerabilities and develop appropriate countermeasures for attack detection, it is preferable to emulate rather than simulate the communication networks.

### Related Work

There are various approaches in current research that depict SGs using both hardware within laboratories and software in the form of simulations. In the field of co-simulation for power grids and ICT, various approaches already exist that usually pursue a specific investigation objective in terms of the impact of cyberattacks, ICT malfunctions, or ICT infrastructures for monitoring and control applications [5]. In this work, we understand Digital Twin (DT) to be a system that emulates a physical target system through continuously modeling-based simulation of the physical target system's functionalities [6]. In particular, we additionally understand CPDT being a DT that involves "the exchange of data between the replica and the target physical system, based on which the physical state of the system and its composition can be determined and updated" [11]. These capabilities can therefore be used "to predict, control, and optimize the functionalities of the target physical system while interacting in the form of feedback to adapt to environmental changes" [11]. Several use cases are pursued with DT approaches, including a DT approach in advanced metering infrastructures within home networks consisting of smart meters and Internet-of-Things (IoT) devices [7]. Additionally, considering the increasing integration of IoT devices and the associated lifecycle management challenges, DT-based approaches are also becoming interesting for SG lifecycle security assessment [8]. However, these works have only been presented on a conceptual basis. A specific use case that we pursue in this work is to study the cybersecurity of critical infrastructures such as power grids in a secure and controlled environment. In this area, several research works have been conducted to investigate the security status of power grids using cyber-physical system testbeds. Our goal is to develop a co-simulation framework integrated into a laboratory environment with the ability to replicate SG in its process and ICT layers and to simulate cyberattacks by systematically mapping such events for individual components or groups of components.

### Problem Analysis

Malfunctions and cyber threats in process networks, both of IT components and of OT components, can never be completely avoided. Therefore, the possible occurrence of such threats should be investigated, both in terms of evaluating existing networks and their expansion. Studies can form the basis for identifying effective measures with regard to the planning of process networks to make the overall system secure against cyber threats. When analyzing cyber threats with respect to SG cybersecurity, many issues and challenges must be considered. In particular, the issue of missing and insufficient attack data poses a challenge to the validation and testing of security concepts. Our aim is the evaluation of design measures for process networks with respect to the cybersecurity of SGs to enable the evaluation of impact of targeted cyber threats. In this work, a co-simulation approach is used to concurrently simulate the power grid and process network as well as the interrelationships between both systems. An appropriate catalog of cyber threat scenarios will be used in conjunction with the developed co-simulation environment in a cyber-physical laboratory setting to evaluate the replication quality of the CPDT solution in the context of a case study.

## CYBER-PHYSICAL DIGITAL TWIN OF SMART GRID

This section introduces the environment for modeling smart distribution grid use cases. It covers (i) an overview, (ii) integration of operational management concepts, (iii) and replication of multi-stage cyberattacks in CPDT.

### Environment Overview

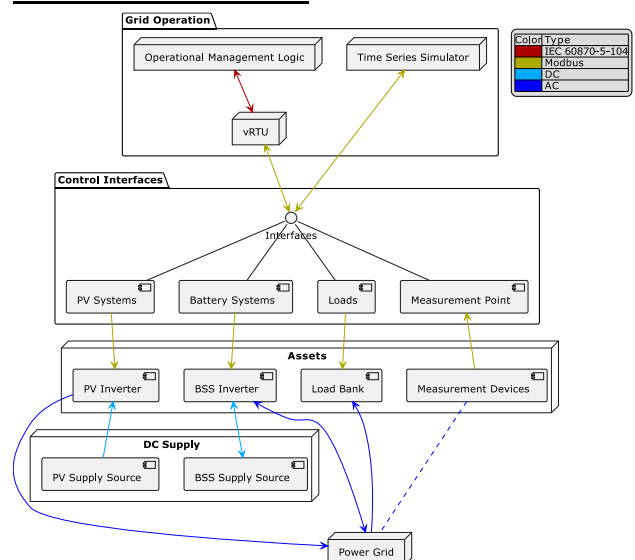

*Figure 1: Overview of CPDT.*

The co-simulation environment is a tool used to evaluate the expansion options for process networks by simulating the power grid and process network as well as their interdependencies. This is achieved by mapping component-specific failures that can occur due to malfunctions or cyberattacks. The goal of the environment is to automate the execution of large-scale scenarios to simulate realistic topology sizes (cf. Figure 1). The co-simulation environment is based on previous work [10, 11] and is built using Pandapower for steady-state simulation of the power grid, and emulated components of the process network are deployed either using a Kubernetes cluster or Docker Containers via Containernet. The step-based coupling of both sides is done via the co-simulation framework Mosaik. The emulation of the process network components allows for validation against real-world ICT components and communication data flows. The co-simulation environment can be used to test the functionality of countermeasures, and communication data can be further processed as packet captures synchronized in time with the results of the power grid simulation and log data of the individual ICT components. This co-simulation environment focuses on attack replication by simulating realistic flexibility coordination processes for small-scale decentralized energy resources.

**Operational Management Concepts**

Coordinating behind-the-meter flexibility, such as heat pumps, electric cars, and battery storage systems, is a major challenge in active distribution grids. The consideration of such use cases requires the development of multi-use strategies. Different stakeholders are interested in using this flexibility, including the asset owner who wants to minimize costs, the DSO who wants to optimize grid operation, and market actors such as Virtual Power Plants. DSOs are responsible for the proper operation of their grid segments, and must ensure that no technical limits are violated. Cyber-resilience is essential for these systems as they require a high degree of networking. We use the concept of Virtual Edge Devices (VEDs) for modeling smart homes [9]. The VEDs are a flexible framework for simulating and emulating Energy Management Systems (EMS). The framework abstracts the information inputs and outputs through interfaces, which can be connected to other simulators or real-world components. The core of a VED is a centralized key-value store that ensures data consistency across various logics and interfaces, and realizes the abstraction between different components. The VED concept also allows for the integration of external systems, such as web-based flexibility coordination platforms, through interfaces. This enables the flexible use of simulated or real components on all layers, and enables the analysis of the cyber-resilience of different IT-architectures. The process network in active distribution grids is simulated using a virtual Remote Terminal Unit (vRTU) developed within this work. This vRTU is a modular container that supports industrial communication protocols such as IEC 60870-5-104 and Modbus, a Python-based logical interface, and additional services such as SSH, SNMP, FTP, Telnet, and HTTPS. It also allows for the integration of cyber vulnerabilities such as remote code execution and privilege escalation. The vRTU is used to link various data points and simulate the behavior of physical RTUs and IEDs. This approach enables the lateral propagation of cyberattacks and allows for the use of the vRTU in both co-simulation and software-in-the-loop in the laboratory environment. Additionally, components of the laboratory environment can also be used in a hardware-in-the-loop approach or other external systems can be integrated for further development and validation purposes.

**Cyber-Physical Digital Twin**

To replicate a reference grid in a CPDT, all devices and their network traffic must be modeled. Physical devices are represented by Docker containers, while network devices like switches and firewalls are modeled through code. The Human Machine Interface (HMI), MTU, vRTU, and IED components are all coded to simulate their physical counterparts. These components are all deployed in different Docker containers, with the HMI and MTU in the same container. The network between the components is emulated using Containernet, which allows for the use of Docker containers as hosts. This enables the real-time observation and capturing of traffic passing through interfaces, similar to the laboratory. The MTU and vRTU communicate using the IEC60870-5-104 protocol, while the vRTU and IED communicate using the Modbus protocol. The power grid is simulated using Pandapower.

**Cyberattack Replication**

The first step in simulating cyberattacks on the power grid is to identify specific attack vectors that could impact the reliability of the grid. One example of such a scenario is a machine-in-the-middle (MITM) attack on the communication infrastructure. This involves the attacker intercepting communications between virtual MTUs and RTUs, and manipulating commands sent between them, gaining control over their behavior. To test the effectiveness of such an attack, a normal scenario is first established, simulating grid operations without any attack effects. The attacker's commands are then sent through the legitimate MTU, to demonstrate that they can take over control of the control units through unprotected communication paths. This helps to evaluate the potential success of external unauthorized third parties. It also highlights the potential for undetected manipulation of control commands and measured values, and the potential for experts to manipulate the readings to appear valid to the MTU. Due to the lack of security mechanisms implemented in the practical production environment, such attack vectors can be enabled by unauthorized third parties. Another example is an attack that starts at the operating level on a workstation, manipulating update files and uploading them to the file exchange server, then manipulating the code that transmits commands to the MTU. This can also cause the vRTU to slow down or crash. The attacker can also initiate an attack on the internal network, performing a network scan and attempting to gain access to the device using default passwords. If successful, the attacker can execute arbitrary commands on the device, resulting in the same effects as the first scenario. This scenario also allows for other actions, such as reading critical data, to be observed on the network.

## CASE STUDY INVESTIGATION

We evaluate the proposed environment through a case study and discuss the quality of replicated scenarios for security research, based on the previously introduced environment.

### Laboratory Setup

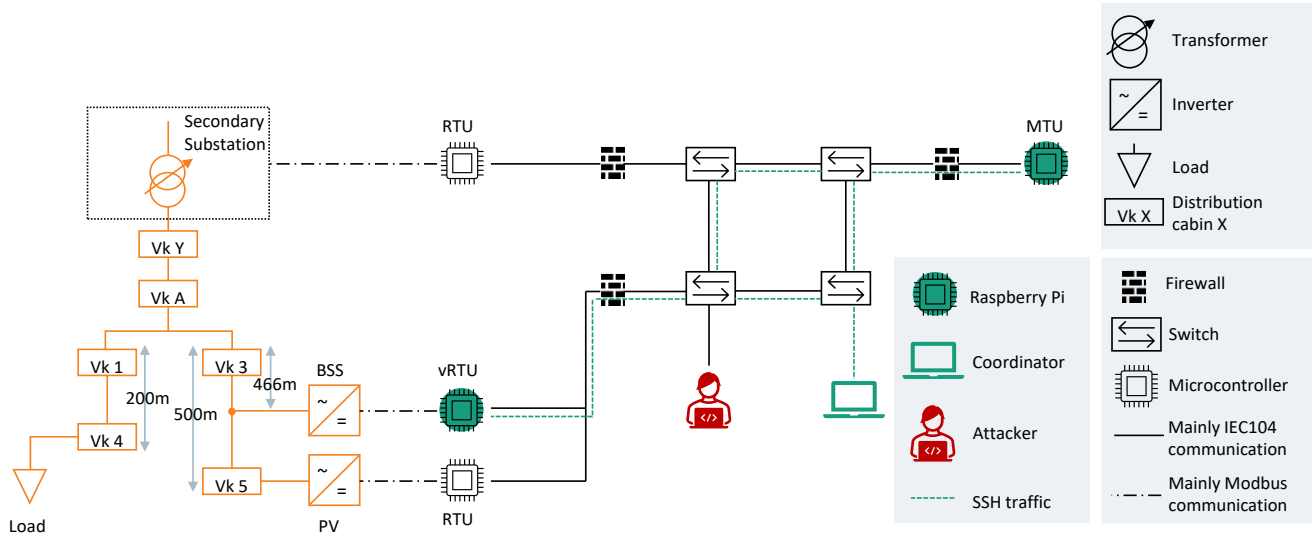

*Figure 2: Illustration of the cyber-physical laboratory environment depicting the reference grid.*

The laboratory setup shown in Figure 2 was used to simulate cyberattacks on power grids [4]. It consists of a medium- and low-voltage grid with various decentralized generation plants, connected to the process data network and the control center (MTU) via telecontrol devices. The aim of the investigations is to use the generated measurement data to validate the models used. The laboratory setup includes electrical equipment such as a 10 kV / 0.4 kV controllable transformer, a battery storage system (BSS), and photovoltaic (PV) inverters. The low voltage network consists of two cable strands and is equipped with integrated current and voltage measurement points. The distributed generation systems are commercial inverters with adjustable feed-in profiles, controlled and monitored via telecontrol devices with Modbus. The process data network of the laboratory setup consists of four ICT switches in ring topology, and the control room is connected directly to the process data network via a firewall. This setup allows for a realistic simulation of primary and secondary components, enabling cross-domain investigation of cyberattack scenarios in power grids and their resulting interactions.

### Experiments

Attacks are first tested on a CPDT before being transferred to a laboratory environment. To prevent damage to real hardware, the HMI, MTU, and vRTU code are installed on Raspberry Pis. The experiment includes three different scenarios: (i) normal operation, (ii) a DoS attack on the vRTU, (iii) and a targeted attack on the vRTU manipulating control commands (cmd) to BSS inverter (increasing charging power). The attacks start at around 200 seconds.

Figure 3 shows the results of experiments conducted in the laboratory testbed. The reference scenario illustrates self-consumption optimization, where local power demands are met by local DERs, reducing the need for external power (power balancing observable on secondary substation power curve). The attack scenarios, specifically the DoS attack, showed the impact of the attack on the vRTU and power imbalance on the secondary substation, and a similar impact can be observed in the targeted cmd manipulation attack. Figure 4 shows the replication of the scenario in the CPDT, with accurate replication of patterns. The replicated patterns allow for the identification of different leveraged attack strategies, indicating a high level of accuracy in the replication. However, some dynamics may not be replicated completely such as power spikes due to the steady-state power simulation. Overall, the results suggest a high level of accuracy in replicating the operation and impact of attacks on the grid.

### Discussion

The experiment in which cyberattacks on a power grid are simulated in a laboratory setting using a CPDT demonstrate the capability of generating complex cyberattack scenarios with severe impact on the grid operation. The CPDT has successfully replicated the attacks in the laboratory setting with minimal challenges, which is a positive indication of the feasibility and accuracy of replicating such attacks in real-world scenarios. The experiment indicates that the CPDT approach can be used to identify areas for improvement in terms of data quality and diversity of attack scenarios.

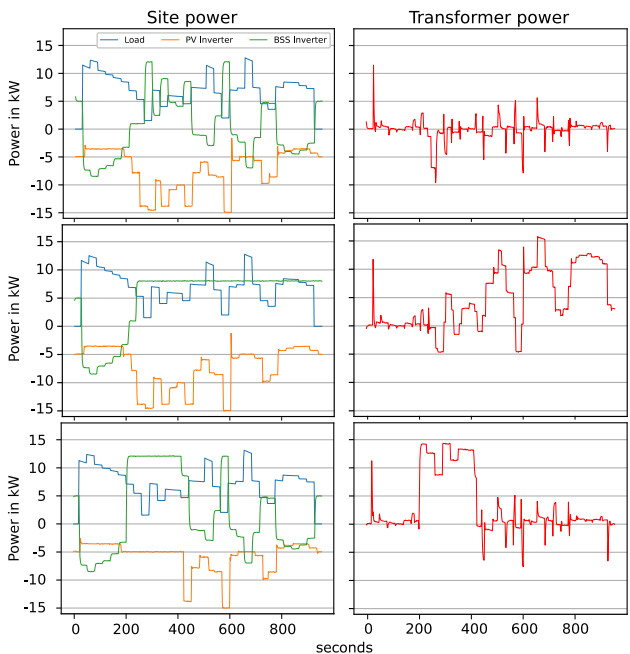

*Figure 3: Testbed results: 1.row no attack, 2.row DoS, 3. row cmd manipulation.*

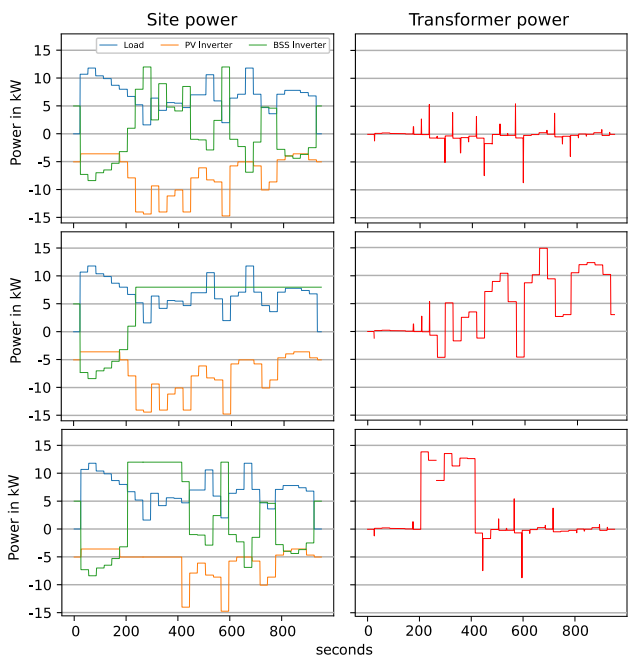

*Figure 4: Results of scenario replication in CPDT.*

In particular, the obtained data can be used for training the machine learning based anomaly detection used to detect attacks. This provides an opportunity for future research to focus on collecting high-quality and representative data for training, which can be extended based on the results of the experiment. Currently, the experiment is not considering the initial stages of the attack, such as the reconnaissance and initial entry into the system. This presents an opportunity for future research to expand the scope of the experiment to include these stages and provide a more comprehensive understanding of cyber-attacks on power grids. Overall, the experiment successfully replicated cyber-attacks on a power grid in a laboratory setting using the presented CPDT approach, and provides a foundation for future research to improve upon in terms of data quality and diversity of attack scenarios. This will help in understanding the real-world scenarios and can be used to improve the security of power grid systems.

## CONCLUSION

In this paper, we describe the co-simulation environment we developed for studying specific applications such as cybersecurity and SG operation. Our approach is modular and allows for flexible simulations in our laboratory. We demonstrated that the environment can be used effectively for both of the examples given. We also demonstrated how to replicate cyberattacks and manipulate data traffic, which can be useful for creating countermeasures in the future. Additionally, we demonstrated how we can integrate concepts for managing decentralized systems into the environment. This serves as a foundation for further research on multi-use strategies and the cyber-resilience of networked systems.